\documentclass[twocolumn,apl,showpacs,preprintnumbers,amsmath,amssymb,superscriptaddress]{revtex4-1}[11pt]

\usepackage[pdftex]{graphicx}
\usepackage{amsmath}
\usepackage{amsfonts}
\usepackage{amssymb}
\usepackage{color}
\usepackage{float}
\usepackage{color}
\begin{document}
\title{Power and efficiency analysis of a realistic resonant tunneling diode thermoelectric}
\author{Akshay Agarwal}
\author{Bhaskaran Muralidharan}
\email{bm@ee.iitb.ac.in}
\affiliation{Cente of Excellence in Nanoelectronics, Department of Electrical Engineering, Indian Institute of Technology Bombay, Powai, Mumbai-400076, India}
\date{\today}
\medskip
\widetext
\begin{abstract}
Low-dimensional systems with sharp features in the density of states have been proposed as a means to improving the efficiency of thermoelectric devices. Quantum dot systems, which offer the sharpest density of states achievable, however, suffer from low power outputs while bulk (3-D) thermoelectrics, while displaying high power outputs, offer very low efficiencies. Here, we analyze the use of a resonant tunneling diode structure that combines the best of both aspects, that is, density of states distortion with a finite bandwidth due to confinement that aids the efficiency and a large number of current carrying transverse modes that enhances the total power output. We show that this device can achieve a high power output ($\sim 0.3$ MW$/m^2$) at efficiencies of $\sim 40\%$ of the Carnot efficiency due to the contribution from these transverse momentum states at a finite bandwidth of $kT/2$. We then provide a detailed analysis of the physics of charge and heat transport with insights on parasitic currents that reduce the efficiency. Finally, a comparison between the resonant tunneling diode and a quantum dot device with comparable bandwidth reveals that a similar performance requires ultra-dense areal quantum dot densities of $\sim 10^{12}/cm^2$.
\end{abstract}
\maketitle
\begin{figure}
	\centering
		\includegraphics[width=2.5in,height=3.0in]{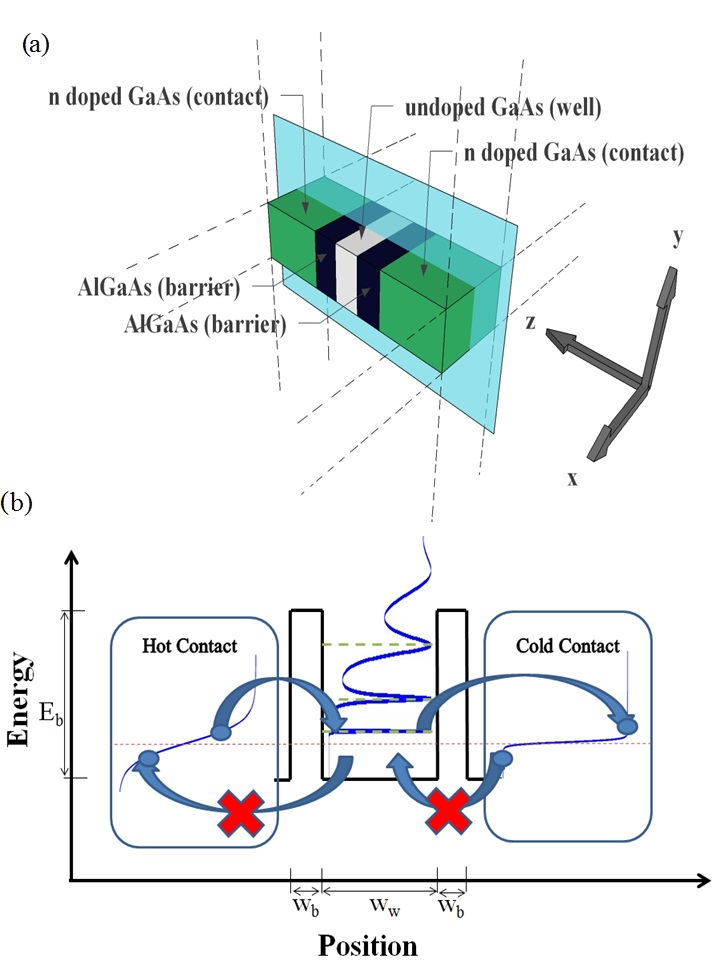}
		\caption{(a) Schematic of the RTD thermoelectric considered here. Two n-doped GaAs regions form the contacts, and a thin GaAs layer (white) between two AlGaAs barriers (blue) leads to energy quantization in the z-direction. Note that the device extends to infinity along both the x and y directions. (b) Conduction band diagram of the RTD heterostructure along the light blue plane. The heterostructure is characterized by parameters $E_b$, w$_w$ and $w_b$. Also shown is the basic thermoelectric operation; the sharp transmission peak allows unidirectional flow of electrons.}
		\label{Fig1}
		\end{figure}
\begin{figure}
	\centering
		\includegraphics[width=2.2in,height=2.8in]{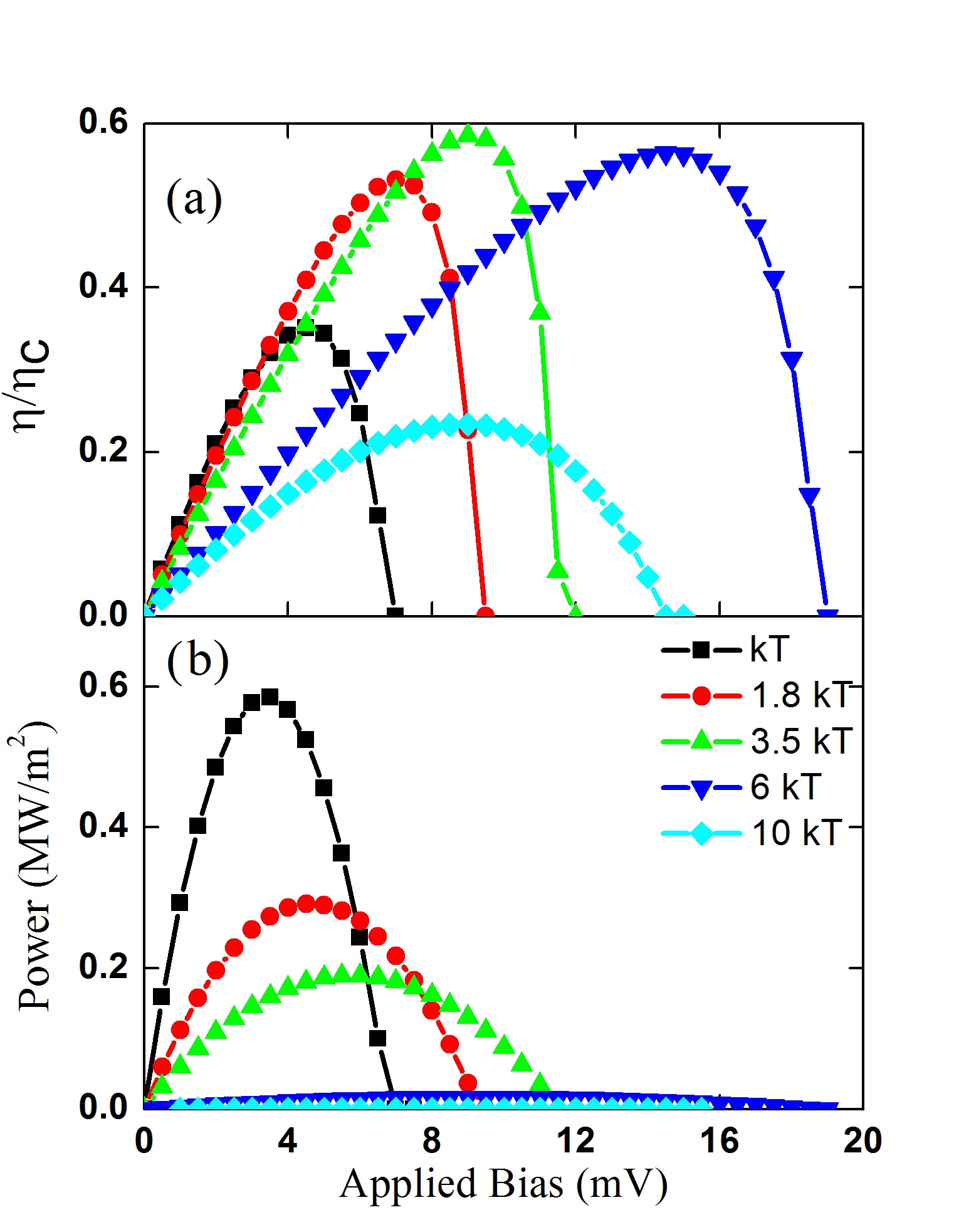}
		\caption{(a) $\eta$ (as a fraction of $\eta_{C}$ ) and (b) Power as a function of applied bias ($V_{applied}$) for various values of $E_{pos}$  (indicated in the legend). The RTD device reaches a maximum efficiency of $60\%$ of $\eta_C$ for $E_{pos}=3.5 kT$ and a maximum power of $0.6$ MW$/m^2$ for $E_{pos}=kT$. Importantly, for $E_{pos}=1.8 kT$  high power ($0.3$ MW$/m^2$) at high efficiency ($\sim 40\%$ of $\eta_C$) is attained.}
\label{Fig3}
\end{figure}
\begin{figure}
	\centering
		\includegraphics[width=2.3in,height=2.8in]{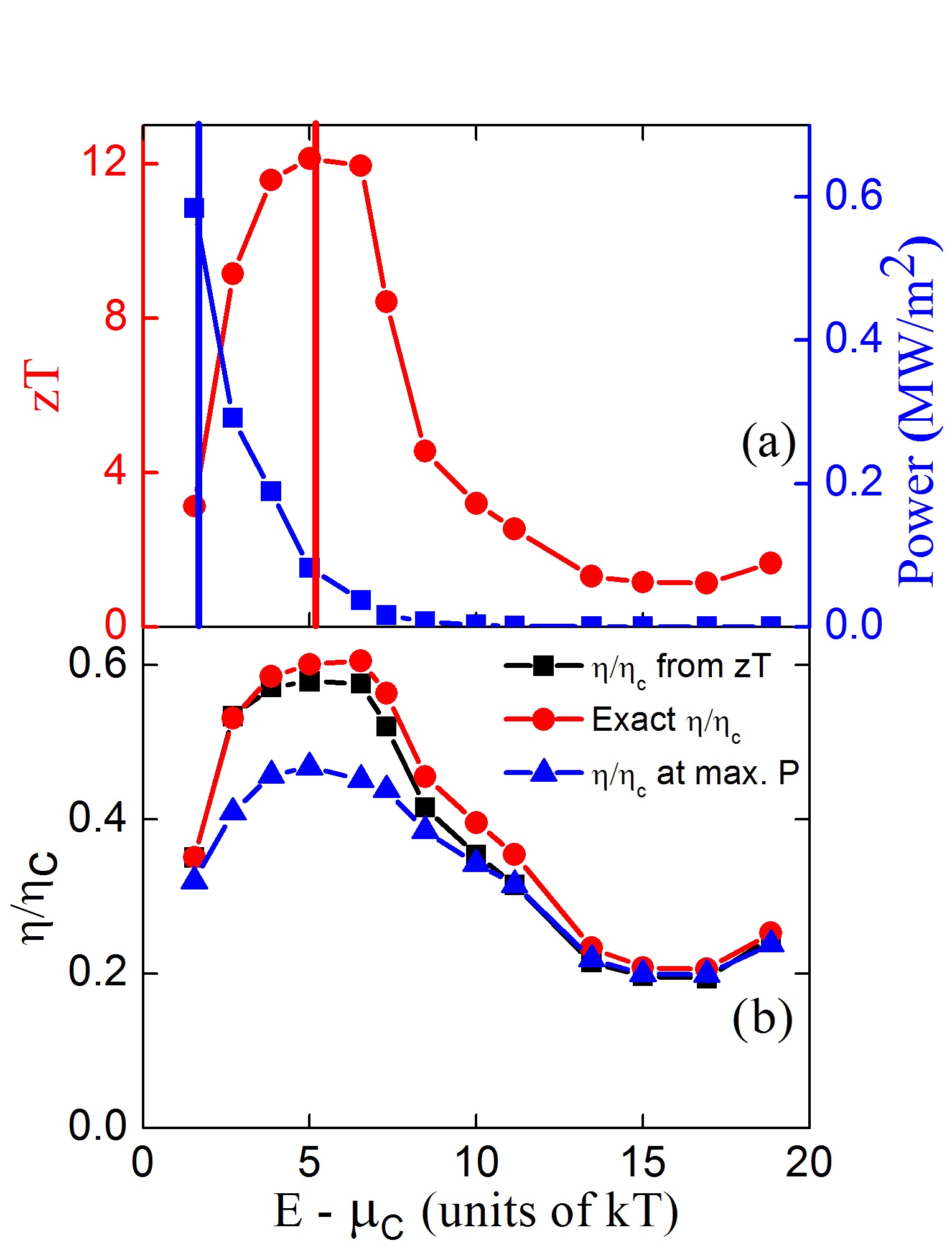}
		\caption{(a) $zT$ and maximum power as a function of ground $E_{pos}$. It can be seen that the maximum $zT$ (12, at $E_{pos} = 3.5 kT$ ) and maximum power ($0.6 MW/m^2$, at $E_{pos} = kT$) are attained at very different points, thus highlighting the power-efficiency tradeoff. However, it is possible to obtain high power at high $zT$, as is evident at $E_{pos} = 1.8 kT$ ($zT=9$, power$=0.3 MW/m^2$).
(b) Efficiency derived from $zT$, calculated maximum efficiency and efficiency at maximum power ($\eta_{max P}$)  as functions of  $E_{pos}$ . $zT$ accurately predicts the maximum efficiency but overestimates $\eta_{max P}$ in the region where power output is high.}
\label{Fig4}
\end{figure}
\begin{figure}
	\centering
		\includegraphics[width=3.5in,height=3in]{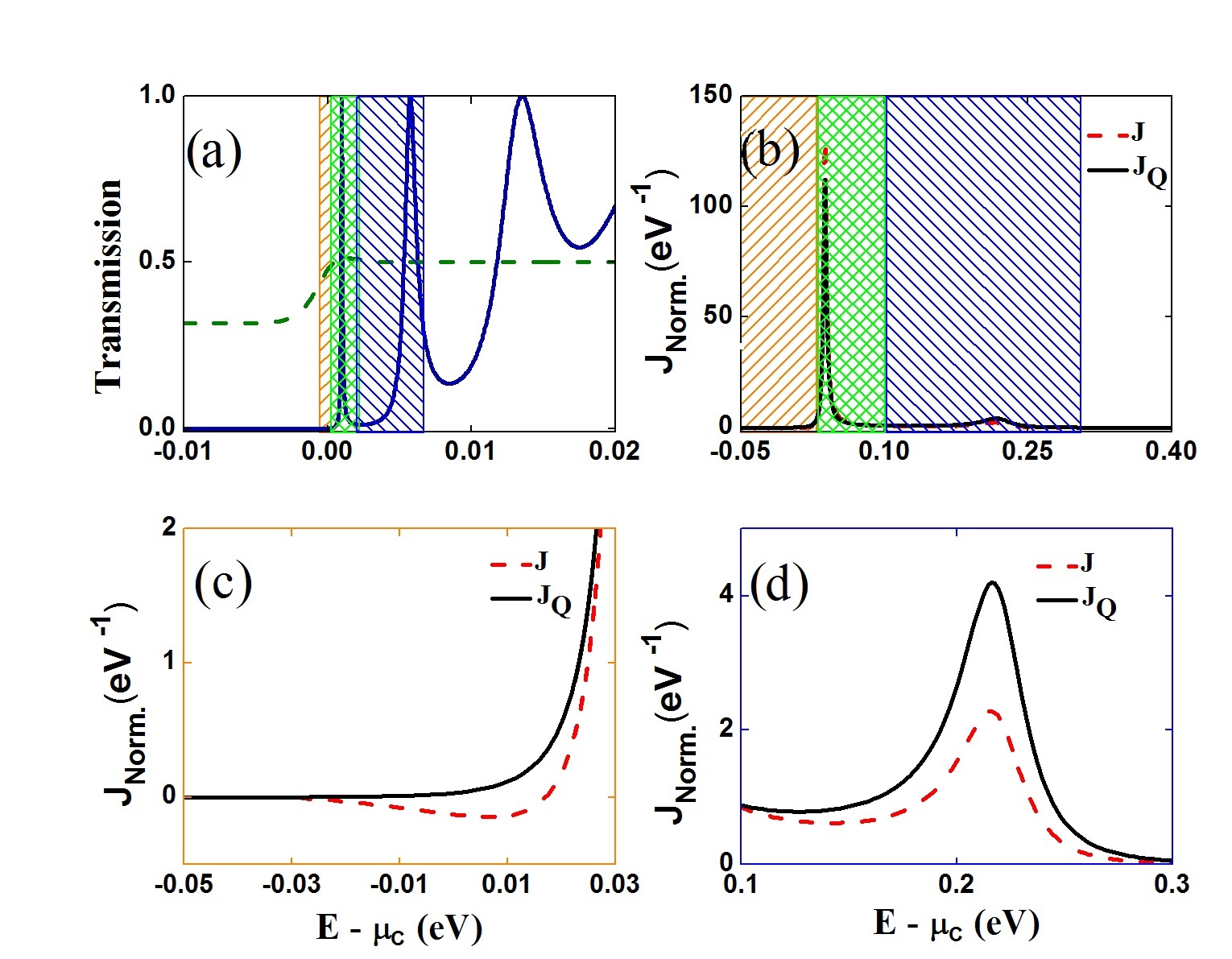}
		\caption{(a)Transmission coefficient and (b) Charge (black solid line) and heat (red dashed line) current densities through the RTD as a function of $E_{pos}$. Also shown as an eye-guide is the difference $F_H-F_C$ displaced by a constant; current is determined by the overlap between the two functions. Shaded regions are the three ``transport windows'' into which electrons are classified  (refer text): low energy (orange), intermediate energy (green) and high energy (blue). (c) Zoomed-in view of the low energy window; while the charge current is negative, the heat current is positive (d) Zoomed-in view of the high energy window; this part of the spectrum makes a larger contribution to the heat than to the charge current. Regions (c) and (d) together restrict the efficiency.}
\label{Fig5}		
\end{figure}

 The thermoelectric figure of merit, $zT$, has traditionally been used to evaluate the performance across various thermoelectrics and is defined as:  
\begin{equation}
zT=\frac{S^2\sigma}{\kappa_{el}+\kappa_{ph}} T,
\end{equation}
where $S$ is the thermopower (Seebeck coefficient), $\sigma$ is the electrical conductivity and $\kappa_{el}$ and $\kappa_{ph}$  are the electronic and lattice (phonon) thermal conductivities respectively. Efforts to improve the thermoelectric performance have focused on reducing $\kappa_{ph}$  through nanostructuring several interfaces in the device \cite{sny08,har02,Chen_2}, or improving the power factor $S^2\sigma$ by modifying the electronic density-of-states (DOS) \cite{dress1,dress2,chen_1,sofo,heremans,nakh}. 
 \indent Following the original proposal of Hicks and Dresselhaus on $zT$ enhancement in quantum well (QW) and wire heterostructures \cite{dress1, dress2}, Sofo and Mahan \cite{sofo} proposed that the optimum ``transport distribution'' (which is related to the the density of states in the device \cite{kim}) for thermoelectric performance was a delta function, achievable in an ideal quantum dot (QD). It was subsequently shown that with a delta-like transport distribution, thermoelectric operation proceeds reversibly at Carnot efficiency $\eta_{C}$ under open circuit (Seebeck) voltage conditions\cite{linke,bhaskaran}. Reversible operation however necessitates {\it{zero power output}}, while finite power output occurs at efficiencies lower than $\eta_{C}$ \cite{bhaskaran}. The figure of merit $zT$, hence, should not be used as a sole indicator of thermoelectric performance specifically if the operation at optimum power is to be considered \cite{nakh, bhaskaran,Espo_1,Espo_2,Espo_3}.
\\ \indent This power-efficiency tradeoff was further elucidated in subsequent works \cite{nakh,bhaskaran} and specifically elucidated in Fig.4 of [9], which clearly demonstrates the disparity between the maximum efficiency $\eta_{max}$ and the efficiency at maximum power $\eta_{maxP}$ for low broadening (low power) and the severe degradation in the efficiency for high broadening (high power). This work \cite{nakh} also demonstrated the possibility of very high (upto $2$ MW$/m^2$) power outputs for 3-D thermoelectrics. Also, Kim et al., \cite{kim} pointed out the high packing fraction and low size requirements for quantum wells and wires in order to realize their potential benefits over bulk thermoelectrics.
\\ \indent Recently, a thermoelectric generator \cite{sanchez1} based on a hot central cavity coupled to two cold contacts through resonant tunneling diode (RTD) heterostructures was proposed. These devices can potentially combine the benefits of a sharp DOS and high power output due to transport through perpendicular momentum modes. Given the device concept that was developed in the aforesaid work, it is important to dwell into the details of the transmission spectra and the effect of excited levels that accompany a realistic RTD structure. 
\\ \indent In this letter, we present a quantitative study of the performance of an RTD-based thermoelectric (Fig.~\ref{Fig1}(a)) with a realistic transmission line width $ kT/2$ of the ground state energy level. We show that the power-efficiency tradeoff is not as severe in this device as in QD devices, and it is possible to obtain high power (upto $0.3$ MW$/m^2$) at an efficiency of $40\%$ of $\eta_{C}$ through it. We also present a detailed analysis of the physics of charge and heat transport in the RTD devices and quantify the effect of high energy resonances and parasitic reverse currents on power output and efficiency. Finally, we compare RTD and QD devices and conclude that a very high ($\sim10^{12} /cm^2$) QD density is needed to match the performance of RTDs. Moreover, the two devices show similar maximum efficiencies, which we explain in terms of the relative widths of the energy levels in the transport window.
 \\ \indent A schematic of the heterostructure we have used in our simulations is shown in Fig.~\ref{Fig1}(a).  The simulated device extends to infinity in the x- and y-directions as indicated by the dotted lines in Fig.~\ref{Fig1}(a). Quantization in the z-direction is achieved by sandwiching a layer of GaAs (white) between two AlGaAs barriers (dark blue). We choose the AlGaAs/GaAs system because the lattice constant shows very little variation over all compositions of AlGaAs and hence modifications to the device bandstructure due to strain are minimal \cite{streetman}. We can thus model these devices fairly accurately using a simple tight-binding Hamiltonian within the one-band effective mass model \cite{datta}. The conduction band diagram of this heterostructure along the light blue plane is shown in Fig.~\ref{Fig1}(b), which also portrays the basic thermoelectric operation at zero bias. Energy filtering will improve as the width of the conducting state reduces, which explains the increase in efficiency with reducing level width. \\
\indent In order to investigate the device properties at various positions of the ground state energy level relative to the equilibrium Fermi level($E_{pos}=E-\mu_{c}$), we vary three parameters: the width ($w_b$) and the height ($E_b$) of the AlGaAs barriers (which can be experimentally realized by varying the relative proportions of aluminium and gallium) and the width ($w_w$) of the GaAs well such that the level broadening is fixed at $kT/2$, where $T=(T_H+T_C)/2$. Previous studies of nanoscale thermoelectrics have primarily focused on very sharp quantized levels to maximize efficiency \cite{nakh, sanchez1, sanchez2} .However a mean-field analysis such as that employed here and previously is only accurate in the limit of large coupling to device contacts \cite{Basky_Beenakker}, which inevitably leads to a large level broadening. Also experimentally grown III-V QD spectra typically show a linewidth $> 10$ meV at room temperature \cite{subho1,subho2}. $kT/2$ is thus a realistic lower bound on level broadenings. Further, a previous study \cite{jeong} considered the optimal bandstructure for thermoelectric performance and found that when the lattice conductivity is taken into account, a broadened dispersion produces a higher $zT$ than an ideal Sofo-Mahan delta function \cite{jeong}. It is thus instructive to study the thermoelectronic performance of finitely broadened energy levels, even though in the present study we have ignored the lattice thermal conductivity. \\
\indent We employ the self-consistent,ballistic Non-Equilibrium Green's Function (NEGF)-Poisson formalism to calculate the transmission at various energies through the device \cite{datta, lake}. In order to apply a bias ($V_{bias}$) across the device, we change the Fermi level of the hot contact. The self-consistent calculation accounts for the non-equilibrium shift in the device  transmission function. The calculated transmission is then used in the Landauer equations for charge and heat current densities \cite{sanchez1}:
\begin{equation}
J=\frac{em^*}{2\pi^2\hbar^2}\int dE_{\perp} dE_z T(E_z)[f_H(E_z + E_{\perp}) - f_C(E_z + E_{\perp})]
\end{equation}
and
\begin{eqnarray}
J^Q_H = \frac{m^*}{2\pi^2\hbar^2}\int dE_{\perp} dE_z (E_z + E_{\perp}-\mu_{H}) T(E_z) \nonumber \\
\quad \left [f_H(E_z + E_{\perp}) - f_C(E_z + E_{\perp})\right ]
\end{eqnarray}
The integration along the transverse co-ordinate is performed assuming periodic boundary conditions along these directions. The equations simplify to:
\begin{equation}
J=\frac{em^*}{2\pi^2\hbar^2}\int dE_z T(E_z)[F_H(E_z) - F_C(E_z)]
\end{equation}
and $J^Q_H=J^{Q1}_H+J^{Q2}_H$, where $J^{Q1}_H$ and $J^{Q2}_H$ are given by:
\begin{eqnarray}
J^{Q1}_H = \frac{m^*}{2\pi^2\hbar^2}\int dE_z (E_z) T(E_z) \nonumber\\
\quad \left[F_H(E_z) - F_C(E_z) \right] 
\label{heat1}
\end{eqnarray}
\begin{equation}
J^{Q2}_H =\frac{m^*}{2\pi^2\hbar^2}\int dE_z T(E_z)\left[G_H(E_z) - G_C(E_z)\right]\\
\label{heat2}
\end{equation}
\\
Here $F_i=\int_0 ^\infty dt (1+e^{t-x})^{-1} = log(1+e^x)$ and $ G_i=\int_0 ^\infty dt \hspace{2pt}t(1+e^{t-x})^{-1} $, with $i=C/H$. The formalism described above enables us to study the energy distribution of the charge and heat currents, and hence characterize the parasitic components of current and secondary resonances that bring down the efficiency. \\
We present the efficiency in Fig.~\ref{Fig3}(a) and the power Fig.~\ref{Fig3}(b) from the device at various values of $E_{pos}$.
We note that a power of $0.6$ MW$/m^2$ at an efficiency of $\sim 32\%$ is obtained for $E_{pos}=kT$, and at $E_{pos} = 3.5 kT$ the maximum effficiency touches $60 \%$ of $\eta_C$. Optimal performance, however, is obtained at $E_{pos} = 1.5 kT$, with a power of $0.3$ MW$/m^2$ at  $40\%$ of $\eta_C$. The figures for maximum power and efficiency we have obtained are similar to the thermionic power generators analysed in \cite{nakh}, the reasons for which will be explored shortly. Here instead we stress on our calculation of the transmission spectrum of the device self-consistently through the NEGF formalism as outlined earlier, instead of inserting it manually as in previous studies. We thus avoid any unrealistic assumptions about the ground state position and width. \\
\indent We have also calculated $zT$ for the RTD device, which is plotted along with the maximum power as a function of $E_{pos}$ in Fig.~\ref{Fig4}(a). We immediately see that $zT$ and maximimum power occur at very different $E_{pos}$, which supports the conclusions of several earlier studies that $zT$ is not a good indicator of the power performance of a thermoelectric. Corresponding to the maximum efficiency point at $E_{pos} = 3.5 kT$ we get a $zT$ of $12$. At the previously mentioned optimal $E_{pos}$ ($1.5 kT$), $zT$ is $9$. It is thus possible to obtain both a high $zT$ and a high power, and thus these devices present, what we believe, a better power-efficiency tradeoff than both QD and bulk thermionic devices by combining the advantages of both. \\
\indent In Fig.~\ref{Fig4}(b) we plot the efficiency calculated from $zT$, the calculated maximum efficiency and the calculated efficiency at maximum power ($\eta_{max P}$) as functions of $E_{pos}$. It is apparent that while $zT$ predicts the maximum efficiency quite accurately over the entire range of $E_{pos}$, in the region of significant power output, however, it overestimates $\eta_{max P}$. This points once again to the unsuitability of $zT$ as the sole design parameter for low-dimensional thermoelectrics. \\
\indent To better understand the factors responsible for limiting the efficiency, we analyze the physics of transport through the thermoelectric device in Fig.~\ref{Fig5}. $J_{norm}$ in Fig.~\ref{Fig5}(b) denotes the charge (heat) current density that has been normalized to the total charge (heat) current. Although Fig.~\ref{Fig5} is plotted for $E_{pos}=kT$ and $V_{bias}=5$ mV, the discussion that follows is very general. \\
\indent Electrons in the device can be classified into three ``transport windows'' on the basis of their energies shown in the shaded regions in Fig.~\ref{Fig5}(a) and (b)): low energy, moving from the cold to hot contact; intermediate energy, moving from hot to cold contact and high energy, also moving from hot to cold contact. Fig.~\ref{Fig5}(c) is a zoomed in version of the low energy window. Charge current density here is negative but the heat current is positive. We similarly zoom into the high energy window in Fig.~\ref{Fig5}(d). Due to their higher energy these electrons make a higher fractional contribution to heat current than the charge current \cite{nakh}. Both these regions together contribute in reducing the device efficiency. \\
\indent As an interesting aside, the presence of secondary resonances in the high energy window of the transmission spectrum (Fig.~\ref{Fig5}(a)) enhances the contribution of this window to the charge and heat currents (Fig.~\ref{Fig5}(c)). Such secondary resonances will be present in a realistic low-dimensional thermoelectric device, but their contribution to degrading the efficiency has not been considered previously. Here, for example, if the secondary resonance is ignored, we find that the efficiency increases from $40\%$ to $46\%$ of $\eta_{C}$. A quantitatively accurate model of low-dimensional thermoelectric device performance must therefore take the entire transmission spectrum and not just a simple Lorentz-broadened level into account. \\
 Unlike QDs, RTD devices do not approach Carnot efficiency even close to the open circuit voltage, because of transport through transverse momentum states. This is easily seen from \eqref{heat1} and \eqref{heat2}. The second term in \eqref{heat2} is not present in QDs. Even if we assume an ideally sharp delta-like energy level, which leads to the Carnot efficiency at the Seebeck voltage $V_{s}$, we see that the second term will lower the efficiency for RTDs.\\
\begin{figure}
	\centering
		\includegraphics[width=2.3in,height=2.8in]{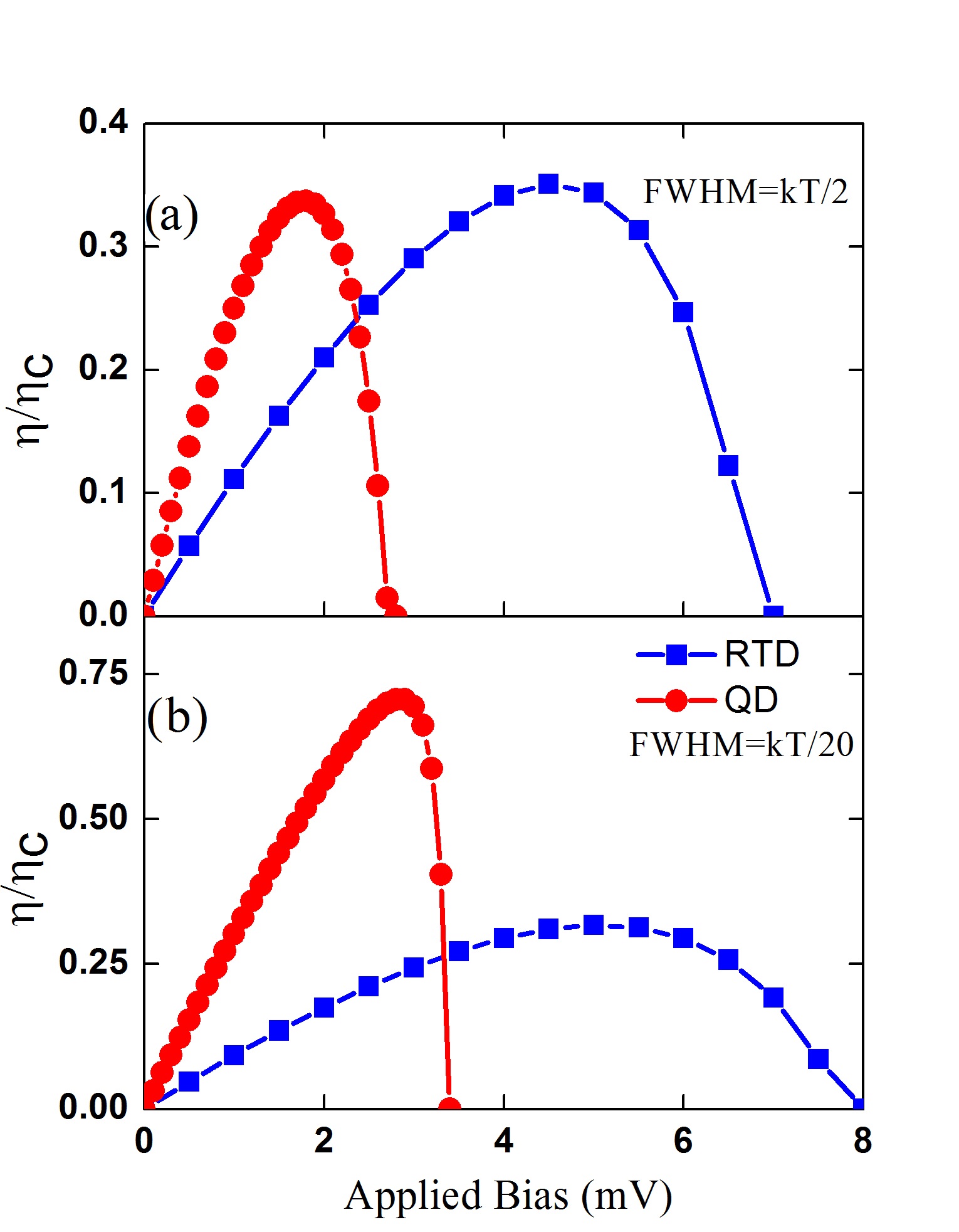}
		\caption{QD and RTD efficiencies as functions of applied bias $V_a$ with level width of (a) $kT/2$ and b)$kT/20$ . While level broadening degrades QD energy filtering significantly, it has a much smaller effect on the RTD.} 
\label{Fig6}
\end{figure}
\indent The advantage of RTD thermoelectrics however lies in the large power output. From \cite{nakh} we see that at resonance widths of $\sim kT/2$, a QD can give a thermoelectric power output of $\sim 30$ pW per dot. To match the power density of an RTD ($0.3$ MW$/m^2$), a QD density of $\sim 10^{12} /cm^2$ is required \cite{kim}. Self-assembled III-V QDs typically display a density of $\sim 10^{10}-10^{11} /cm^2$. Of course, the broadening present in self-assembled QD samples is due to size distribution and hence fundamentally different from the contact coupling-induced broadening discussed here. From \cite{sanchez2} we see that the thermoelectric power output of QDs with sharp levels does not degrade much at disorder-induced linewidths of $\sim kT/2$. Here, however, the power output is really low to begin with, and hence an even higher QD density would be needed to match RTDs. It is also worth noting that the RTD device features a higher Seebeck voltage $V_{s}$ due to the contribution from transverse current carrying modes. 
\\
 \indent Interestingly, we see in Fig.~\ref{Fig6}(a) that the efficiency at maximum power is nearly the same for both the RTD and QD at linewidth $\sim kT/2$. Although Fig.~\ref{Fig6}(a) is for $E_{pos}=kT$, this observation is ubiquitous. While this may seem surprising since a QD is expected to be a much more efficient energy filter than an RTD, we note that both charge and heat transport occur primarily in a window of width $\sim kT$ around the equilibrium Fermi level, where the difference in occupation of the two contacts is significant \cite{sanchez1}. Since the level width is also of order $\sim kT/2$, both the RTD and QD are expected to behave similarly in terms of energy filtering. To confirm this we reduced the width of the dot level to $kT/20$, upon which its maximum efficiency increased to $~60\%$ of $\eta_c$ while the RTD efficiency remained almost unaffected. We thus conclude that not only can the RTD surpass QD thermoelectrics in terms of power output, but can also match its efficiency for realistic ground state energy level broadening. This effect is also responsible for the similarity between the results here and those in \cite{nakh}, who considered a step-like transmission coefficient. \\
\indent In conclusion, we have analyzed the thermoelectric performance of a finitely broadened RTD-based device and shown that it can attain high powers ($0.3$ MW$/m^2$) at high efficiencies ($\sim 40\%$ of $\eta_c$) because of the combined benefits of a large number of transverse momentum modes to carry current and longitudinal energy quantization to enhance filtering. By considering the energy spectrum of the charge and heat currents we estimated the effects of higher energy resonances on the device performance. We also showed that the RTD might be preferable to the QD-based thermoelectric for realistic level broadening. This study however is confined to consideration of the electronic component of thermal conductivity, and a complete understanding of low-dimensional thermoelectrics requires the inclusion of phonon heat transport too. The determination of the best low-dimensional thermoelectric considering both electron and phonon transport, as well as quantification of the power and efficiency performance of this thermoelectric is a fruitful avenue of future research. \\\\
Acknowledgment: This work was supported in part by IIT Bombay SEED grant and the Center of Excellence in Nanoelectronics. The author AA would like to acknowledge useful discussions with Harpreet Arora and Arun Goud Akkala.
\bibliographystyle{apsrev}
\bibliography{ref_TE}
\end{document}